\documentclass[twocolumn,english,aps,pra,floatfix,superscriptaddress]{revtex4}
\usepackage[T1]{fontenc}
\usepackage[latin1]{inputenc}
\usepackage{amsmath}
\usepackage{graphicx}
\usepackage{amssymb}

\makeatletter

\providecommand{\tabularnewline}{\\}

\usepackage{bm}
\usepackage{graphicx}
\usepackage{epsfig}
\usepackage{latexsym}
\usepackage{color}
\usepackage{subfigure}
\usepackage{array}

\newcommand{\ket}[1]{|#1\rangle}

\usepackage{babel}
\makeatother
\begin{document}

\title{Global control and fast solid-state donor electron spin quantum
computing}

\author{C. D. Hill}
\email{hillcd@physics.uq.edu.au}
\affiliation{Centre for Quantum Computer
Technology, and Department of Physics, The University of Queensland, St
Lucia, QLD 4072, Australia}

\author{L. C. L. Hollenberg}
\affiliation{Center for Quantum Computer Technology,
School of Physics, The University of Melbourne, Victoria 3010,
Australia}

\author{A. G. Fowler}
\affiliation{Center for Quantum Computer Technology,
School of Physics, The University of Melbourne, Victoria 3010,
Australia}

\author{C. J. Wellard}
\affiliation{Center for Quantum Computer Technology,
School of Physics, The University of Melbourne, Victoria 3010,
Australia}

\author{A. D. Greentree}
\affiliation{Center for Quantum Computer Technology,
School of Physics, The University of Melbourne, Victoria 3010,
Australia}

\author{H.-S. Goan}
\affiliation{Centre for Quantum Computer Technology, University of
New South Wales, Sydney, NSW 2052, Australia}

\begin{abstract}
We propose a scheme for quantum information processing based on donor
electron spins in semiconductors, with an architecture complementary
to the original Kane proposal. We show that a na\"{i}ve implementation
of electron spin qubits provides only modest improvement over the Kane
scheme, however through the introduction of global gate control we are
able to take full advantage of the fast electron evolution
timescales. We estimate that the latent clock speed is 100-1000 times
that of the nuclear spin quantum computer with the ratio
$\mathrm{T_{2}/T_{ops}}$ approaching the $10^{6}$ level.
\end{abstract}

\maketitle

\section{Introduction}

The interest in constructing the components of a solid-state quantum
computer (QC) device where the logical qubits are encoded by single
donor spin\cite{Kan98,VRYE+00} or charge\cite{HDW+04} degrees of
freedom is largely based on the nexus to scalable fabrication
technology in the semiconductor industry. The nuclear spin Kane
QC\cite{Kan98}, is of particular interest due to the relatively long
coherence timescale of P-donor nuclear spins, which bodes well for
qubit storage. On the other hand, simulations of electron exchange
mediated two-qubit logic gates in the Kane scheme\cite{FWH03,HG03,
HG04} showed that the gate fidelity is limited primarily by the
electron coherence where the dephasing timescale was expected to be
closer to the typical gate operation time of O($\mu$s). Recent
measurements\cite{TLA+03} indicate that the coherence time for
phosphorus donor electron spins in silicon is considerably longer -
greater than 60 ms at $T$=4K. This surprisingly long coherence time
means that donor electron-spin based quantum computers may be a more
desirable goal in terms of relative simplicity of qubit
identification, readout, and inherent gate speed.

Proposals for donor-electron spin quantum computing as variations on
the original Kane theme already exist. That of Vrijen et
al\cite{VRYE+00} based on g-factor engineering calls for the
fabrication of complex hetero-structures, and the ability to drag the
electron wave function into high-g regions without ionisation.  The
``digital'' quantum computer concept\cite{SDK03} relies on the ability
to coherently transport electron spins along the Si-oxide interface
using surface gates. The use of electron spins in quantum dot systems
has been considered several times previously, for example in GaAs
systems \cite{LD98} and Si-Ge heterostructures \cite{FRS+02}. A
phosphorous donor electron QC based on the dipole interaction was
proposed in Ref. \cite{dSD04}. A recent review of silicon
quantum-computer architectures can be found in Ref. \cite{HWG}.

\begin{figure}
\centerline{\includegraphics[height=5.5cm]{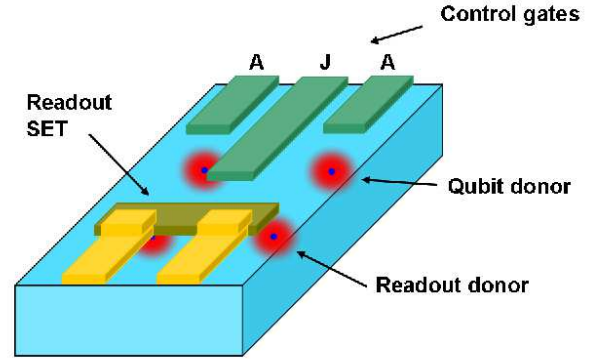}}
\caption{Donor electron spin qubits in the Kane configuration
including A-J-A control gates, auxiliary read-out donors and SET
readout.} \label{fig:qubit}
\end{figure}

\begin{table}
\begin{tabular}{|c|c|c|c|c|c|}
\hline Qubit & $T_{\rm X}$ & $T_2/T_{\rm X}$ & $T_{\rm CNOT}$ &
$T_2/T_{\rm CNOT}$
\tabularnewline \hline
\hline
%
%
n-spin &  6$\mu$s & $10^4$ & 16$\mu$s & $4\times 10^4$
\tabularnewline  &   &  &  &
\tabularnewline \hline e-spin &  2$\mu$s & $3\times 10^4$ & O(10
$\mu$s) & O($10^3$)
\tabularnewline (local control) &   &  &  &
\tabularnewline \hline e-spin &  30ns & $2\times 10^6$ & 148ns &
$6\times 10^5$
\tabularnewline (global control) &   &  &  &

\tabularnewline \hline
\end{tabular}
\caption{Table summarising the relative time-scales for locally
controlled nuclear\cite{HG03} and electron spin qubits compared to
the globally controlled electron spin case. For both nuclear and
electron spin qubits the effective dephasing time is taken to be
the faster of the two, $T_2>60$
ms\cite{TLA+03}.\label{tab:timescales} }
\end{table}

Between the original Kane proposal and these two variants
we present a new proposal for a solid-state quantum computer where the
qubits are also encoded on the spins of Si:P donor electrons, yet
retaining the relative simplicity of the original Kane design. In this
proposal we literally turn the Kane donor based nuclear spin QC
concept inside out and couple it with new ideas for spin readout. The
phosphorus donors now serve to localize the electron spins in space,
and to provide local qubit addressability through the electron-nuclear
hyperfine interaction.  Contrary to the essential and rather complex
role played by the non-logical spins in the Kane proposal -- the
electron spins -- here the nuclear spins are essentially frozen
spectators. The donor electron spin based quantum computer has
potentially an inherently faster clock-speed than the nuclear spin
version due to the much larger magnetic moment. To fully access this
is non-trivial. By introducing new concepts in global control of spin
qubits and correction of spectator evolution, we show by direct
simulation that the inherent speed of the electron spin time scales
can be fully exploited. Single gate operations are achieved with gate
times down to tens of nanoseconds, commensurate with the exchange
based CNOT gate on the order of $150\mathrm{\, ns}$. We estimate that
the electron spin donor QC will have an inherent clock speed around
100 times that of the nuclear spin QC, with $T_{2}/T_{\rm ops}$
approaching $10^{6}$ (see Table \ref{tab:timescales}). A summary of
recent work on single donor electron spin readout\cite{BG03, RBM+04,
GHH+04, HWP+04,GHG04} completes the proposal.

This paper is organised as follows. We introduce the notion of
effective single-spin gate operation through global control
sumplemented by only weak local control, and correction of spin
spectators in the nanosecond temporal arena of fast electron spin
dynamics. We contrast the gate speeds achieved with the relatively slow
canonical single-spin/single-gate control paradigm, where the gate
operation is limited to the microsecond timescale. In the global
control using weak local control and correction paradigm we
demonstrate how \{X, Y, Z, and Hadamard\} single qubit operations, and
the CNOT gate can be carried out, and provide the actual timescales
through numerical simulations.  We then discuss readout, scale-up
issues and quantum error correction.

\section{Single Qubit Rotations}

\subsection{Qubit Definition}

The architecture of the basic donor electron spin qubit with control
gates and a resonant readout mechanism is shown in Fig.
\ref{fig:qubit}. Single phosphorus nuclei play a primary role as the
localizing centres for donor electron spins which encode quantum
information in the canonical fashion as $\ket{0}=\ket{\downarrow}$ and
$\ket{1}=\ket{\uparrow}$. To begin with, we analyze the dynamics in
the effective spin formalism for which the Hamiltonian for the single
qubit system in the absence of a rotating magnetic field is
\begin{equation}
H_{Q}=\mu_{B}B\sigma_{e}^{z}-g_{n}\mu_{n}B\sigma_{n}^{z}+A(V_{A})
\bm{\sigma}_{e}\cdot\bm{\sigma}_{n},\label{eq:HQ}
\end{equation}
where $B$ is the strength of the constant magnetic field, $\sigma^{z}$
is the Pauli Z matrix with subscripts $e$ referring to electrons and
$n$ referring to the nucleus and $A(V_A)$ is the strength of the
hyperfine interaction.

The hyperfine interaction between electron and nucleus is
controlled in the usual Stark-shift manner by varying the bias,
$V_{A}$, on the A-gate in order to deform the electron wave
function $\psi(r,V_{A})$ around the nucleus thereby changing the
hyperfine coupling $A(V_{A})$ as
$A(V_{A})\propto|\psi(0,V_{A})|^{2}$.

\begin{figure}
\centerline{\includegraphics[height=3 cm]{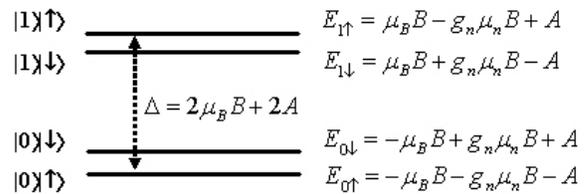}}
\caption{The energy levels of the donor electron-nucleus system in
a magnetic field $B$ and hyperfine coupling $A$. The notation is
$\sigma^z_e = $0,1 (logical qubit states) and $\sigma^z_n =
\uparrow,\downarrow$.} \label{fig_levels}
\end{figure}

It proves beneficial to restrict the Hilbert space of the non-qubit
spin: i.e. the nuclear spin space - in our case the lowest energy
state corresponding to the nuclear spin up. For the Kane nuclear spin
quantum computer the non-qubit electron spins were frozen out by
through the large $B=2\mathrm{{T}}$ background field leading to an
relative electron spin-up/spin down polarization of $10^{-12}$ at
100mK. Here the field serves a similar purpose, but also relies on the
extraordinary long $T_{1}^{(n)}\approx$ 1 hour of the donor nuclear
spin. Since $T_{1}^{(n)}$ is much longer than any other timescale in
the system, the nuclear spin once initialised in the up state (lowest
energy state) is for all intents and purposes predictably inert. The
Kane concept is thus turned inside out.

\subsection{Single Qubit Hamiltonian}

We now describe the canonical method of controlling and manipulate
electron spins. In the following section, section \ref{sec:Xrot} we
will describe how this method may be improved upon. 

By biasing the A-gate correctly we are able to select the qubit system
(the \emph{targetted} qubits). In the canonical method, the A-gate
bias tunes the hyperfine interaction, bringing the qubits into
resonance with background RF field $B_{ac}$ and giving us the ability
to perform single qubit rotations as required. To gain insight into
the canonical control of the electron spin a RF field of frequency
$\omega_{ac}$ we write the single-spin electron Hamiltonian as
(assuming frozen nuclear dynamics in the up state):
\begin{eqnarray}
H_{Q}&=(&\mu_{B}B_z + A(V_A))\mathbf{\sigma_{e}^{z}}\nonumber\\
&&\quad\quad+\mu_{B}B_{ac}(\mathbf{\sigma_{e}^{x}}
\sin\omega_{ac}t+\sigma_{e}^{y}\cos\omega_{ac}t).
\label{eq:HQa}
\end{eqnarray}
This turns out to be a good assumption for typical parameters expected
for the Kane architecture, as we show by numerical simulation
including both nuclei and electrons.

For an initial state $\ket{0}$ the well known Rabi solution gives
the probability of the electron being found in the state $\ket{1}$
after time t as
\begin{equation}
P_{1}(t)=\left(\frac{\mu_{B}B_{ac}}{\Omega}\right)^{2}\sin^{2}\left(\frac{\Omega
t}{\hbar} \right),\label{eq:PRabi}
\end{equation}
where $\Delta\omega = \omega(A) - \omega_{ac}$ is the
detuning between the field quanta and the resonant frequency of
the qubit levels $\omega(A)$, controlled by the hyperfine gate
$A(V_A)$. $\Omega(A)^{2}=(\mu_{B}B_{ac})^{2}+\hbar^{2}(\Delta\omega)^{2}$
and the resonant frequency $\omega(A)$ is given to second order
by
\begin{equation}
\omega(A)=2\left(\mu_{B}B+A+\frac{A^{2}}{\mu_{B}B+g_{n}\mu_{n}B}\right).
\label{eq:Omega}\end{equation}

In the canonical scheme, being able to perform single qubit rotations
is contingent on the ability to shift the electron spin in and out of
resonance with the RF field. Calculations show that by applying a
voltage to the A-gate, one can effectively shift A \cite{KG03}. A
natural state of operation is to tune the frequency of the rotating
magnetic field to the maximally detuned state $\omega_{ac}=\omega(0).$
In the canonical scheme, when no bias is applied to the A-gates,
$A=A_{0}$, and each qubit is out of resonance. When a bias voltage is
applied, the qubits are forced into resonance with the magnetic
field.

In order for the canonical scheme to work, $\Delta\omega$ must be
large compared to the full width half maximum (FWHM) of the resonance
to achieve fidelities at the $10^{-5}$ level. The FWHM is given by
$4\mu_{B}B_{ac}/\hbar$. Clearly to locally control spins using this
method, we must reduce $B_{ac}$ at the expense of gate operation
time. For an error of $P_{1}\approx10^{-5}$ for the off resonance
qubits (ie. those not taking part in the operation) one requires
$B_{ac}\approx10^{-5}\mathrm{{T}}$. This leads to a gate operation
time of $1.7\mu s$ for the qubit being addressed. In Table
\ref{tab:timescales}, this is referred to as the locally controlled
electron spin case.

There are several apparent problems with this canonical single-gate
scheme with an always on AC field. First, the microsecond timescale is
slow compared to the natural Z evolution of the electron spins in the
$B_z=2\mathrm{T}$ field. Also, as a result of this fast evolution, one
must be able to tune the `A' gate at the frequency of the Z evolution
in order to optimise the fidelity of the gates. While this might be
possible, we propose an alternative scheme exploiting global control
which takes full advantage of the fast timescales.

It is considerably simpler to understand the basic control
processes once we transform the single qubit Hamiltonian into a
frame rotating with the RF field. We thus make the substitution
\begin{equation}
\ket{\phi}=\exp\left(\frac{{i\omega_{ac}t}}{2}\sigma_{e}^{z}\right)\ket{\psi},
\label{eq:RotPhi}
\end{equation}
where $\ket{\psi}$ is the wavefunction in the stationary
frame, and $\ket{\phi}$ is the wavefunction in the frame which
rotates at the same frequency as the field $B_{ac}$. The
Hamiltonian in the rotating frame is
\begin{equation}
\tilde{H}_{Q}=\hbar\Delta\omega\sigma_{e}^{z}+\mu_{B}B_{ac}\sigma_{e}^{x},
\label{eq:Hrot}
\end{equation}
The Hamiltonian given in equation (\ref{eq:Hrot}) represents spin
precession or rotation around an axis in the
$\vec{n}=\hbar\Delta\omega\hat{k}+\mu_{B}B_{ac}\hat{i}$ direction.

In order to take full advantage of the fast timescales in the
system we consider an alternative approach for single qubit
rotations to the locally controlled case as we anticipate only
having limited control over $\omega(A)$. In this proposal we
effectively perform single qubit operations by rotating around the
$x$-axis (when $\Delta\omega=0$) and around an axis which is
slightly rotated with respect to this axis described by
$\tilde{H}_{Q}$.

\subsection{X Rotations} \label{sec:Xrot}

In this section we describe the globally controlled qubit operation in
the context of an X rotation, by which we mean a rotation around the
x-axis. To perform an X rotation, we begin with the resonant magnetic
field $B_{ac}$ tuned to the electron resonance obtained when no
voltage is applied to the corresponding A-gate ($A=A_0$), i.e.,
\begin{equation}
\omega_{ac}=\omega(A_{0}).\label{eq:ResonaceCondition}
\end{equation}
In the case when no voltage is applied to an A-gate, electrons will
undergo a rotation around the $x$-axis $\hat{n_{0}}=\hat{\imath}$,
since $\Delta\omega=0$. They will precess with an angular frequency of
$\Omega_{0}=2\mu_{B}B_{ac}$.  This is the natural frequency of
rotation in the system. In the absence of any external influences,
every electron precesses at the same rate.

We now consider how to rotate one of the qubits (the target qubit)
with respect to the others (the spectator qubits). The speed of
rotation of a detuned electron is greater than an electron which
is resonant with $B_{ac}$; that is, $\Omega(A)\ge\Omega_{0}$.
Therefore, if we detune an electron from the resonance, it will
perform a $2\pi$ rotation in less time than every other qubit
requires to do a $2\pi$ rotation around the $\hat{\imath}$ axis.
In fact every other qubit will undergo a rotation of

\begin{eqnarray}
\theta_{x}(A) & = & 2\pi-\frac{2\pi}{\Omega(A)}\Omega_{0}\nonumber \\
 & = &
 2\pi-\frac{2\pi}{\sqrt{(\mu_{B}B_{ac})^{2}+\hbar^{2}(\Delta\omega)^{2}}}\mu_{B}B_{ac}\nonumber\label{eq:ThetaX}
\end{eqnarray}
Since $\Delta\omega$ is constrained, the maximum angle which may be
rotated in a single step is also constrained. By repeatedly applying
this operation and tuning the voltage on the A-gate, an X rotation by
an arbitrary angle $\theta$ may be constructed. It is convenient to
choose $B_{ac}$ such that any rotation up to $\theta=\pi$ may be
performed in a single step. This is possible for typical parameters
when $B_{ac}\le1.2\times10^{-3}\ \mathrm{T}$. After this step, the
target qubit will not be rotated with respect to its original state,
but all the spectator qubits will have undergone a rotation of
$R_{x}(-\theta)$.

The second step required is a correction that rotates every qubit,
both the target and spectator qubits, by $R_{x}(\theta)$. In this step
we bring every qubit in the system into resonance with the magnetic
field, and perform an equal X rotation on each qubit. In the first
step, each spectator qubit was rotated by $R_{x}(-\theta)$. In the
second step everything is rotated by $R_{x}(\theta)$. These angles
cancel and therefore no net operation is performed on the spectator
qubits. The targeted qubit is effectively not rotated at all by the
first step. The second step rotates the targeted qubit by
$R_{x}(\theta)$. The targeted qubit therefore has an overall rotation
of $R_{x}(\theta)$. The steps required for a full X rotation are shown
in Table \ref{tab:XSteps}.

\begin{table}
\begin{tabular}{|c|c|c|c|}
\hline 
Step& Target Qubit& Spectator Qubits& Time (ns)
\tabularnewline \hline \hline 1& $I$& $R_{x}(-\theta)$&
14.8\tabularnewline \hline 2& $R_{x}(\theta)$& $R_{x}(\theta)$&
14.8\tabularnewline \hline Overall& $R_{x}(\theta)$& $I$&
29.7\tabularnewline \hline
\end{tabular}

\caption{Control steps in the single qubit X rotation showing the
operations effected on both target and spectator qubits.
\label{tab:XSteps} }
\end{table}

\begin{figure}
\includegraphics[%
  height=6cm,
  keepaspectratio]{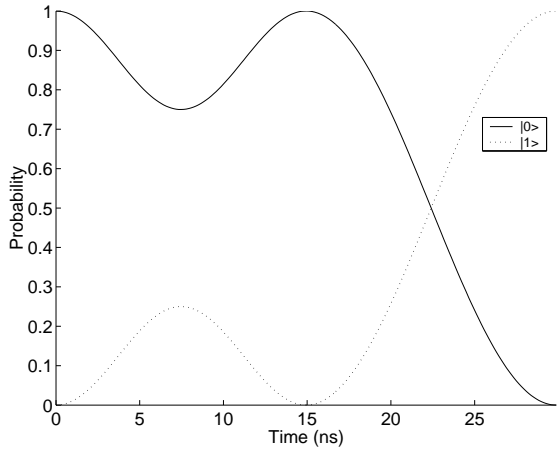}
\caption{Typical X gate evolution and timescales for input states
as indicated. \label{fig:XEvolution}}
\end{figure}

The overall time required for an X gate (i.e. $R_{x}(\pi)$) is
approximately $t_{x}=\mathrm{{29.7\ ns}}$. Often a correction step can
be combined with other correction steps. Not including the correction
step (Step 2), the time required for an X gate is around half this
value at $t_{x}=\mathrm{14.8\ ns}$.  A numerical simulation of this
gate was calculated, and a typical evolution is shown in Fig.
\ref{fig:XEvolution}, using the full Hamiltonian including both
nuclear and electronic spin.

\subsection{Y Rotations}

To achieve a Y rotation, we make use of detuned rotations around the
axis

\begin{equation}
\hat{n}=\cos\phi\hat{\imath}+\sin\phi\hat{k}\label{eq:nAxis}\end{equation}
 Note that we can tune the voltage on an A gate and therefore $A(V_A)$
to produce an arbitrary angle $\phi<\phi_{\max}$. The value of $A$
required may be calculated for an arbitrary angle $\phi$, because we
know that

\begin{equation}
\tan{\phi}=\frac{{\Delta\omega}}{\mu_{B}B_{ac}}\label{eq:Phi}\end{equation}
 This equation allows us to solve for $\Delta\omega$, and therefore
$A(V_A)$. The maximum angle which may be obtained is

\begin{equation}
\tan\phi_{\max}=\frac{\Delta\omega_{\max}}{\mu_{B}B_{ac}}.\label{eq:phiMax}
\end{equation}
Now we consider rotations around this axis,
\begin{eqnarray}
 & & R_{x}(\pi)\ R_{n}(\pi)\ R_{x}(\pi)\ R_{n}(\pi)\nonumber \\ & = &
 (\cos\phi X-\sin\phi Z)(\cos\phi X+\sin\phi Z)\nonumber \\ & = &
 \cos2\phi\ I-i\sin2\phi Y\nonumber \\ & = &
 R_{y}(4\phi)\label{eq:Ry}\end{eqnarray} We may apply $R_{x}(\pi)$
 rotations in parallel on every qubit.  $R_{n}(\pi)$ rotations may be
 applied by detuning the target qubit.  This technique allows for
 arbitrary rotations on the target qubit around the $\hat{\jmath}$
 axis, up to a rotation of $R_{y}(4\phi_{\max})$.  For rotations
 larger than this angle, one may simply repeat the procedure.

After this operation is complete, a correction step may be
required. The target qubit will have undergone a rotation of
$R_{y}(4\phi)$ and all spectators will have been rotated by an angle
$R_{x}(\gamma)$ as they are in resonance with the magnetic field
during the entire operation. Therefore if the total time of the
operation is $t$ then $\gamma=2\mu_{B}Bt$. To correct for this
rotation an additional step is required. We rotate the each
spectator by $R_{x}(-\gamma)$ and effectively do nothing to the target
qubit.  This step is identical to the first step when performing an X
rotation. Each step in this operation is shown in Table
\ref{tab:ySteps}.

\begin{table}
\begin{tabular}{|c|c|c|c|}
\hline Step& Target Qubit& Spectator Qubits& Time
(ns)\tabularnewline \hline \hline 1& $R_{y}(4\phi)$&
$R_{x}(\gamma)$& 50.7\tabularnewline \hline 2& $I$&
$R_{x}(-\gamma)$& 38.3\tabularnewline \hline Overall&
$R_{y}(4\phi)$& $I$& 89.0\tabularnewline \hline
\end{tabular}
\caption{Control steps in the single qubit Y rotation showing the
operations effected on both target and spectator
qubits.\label{tab:ySteps}}
\end{table}

\begin{figure}
\includegraphics[%
  width=7cm,
  keepaspectratio]{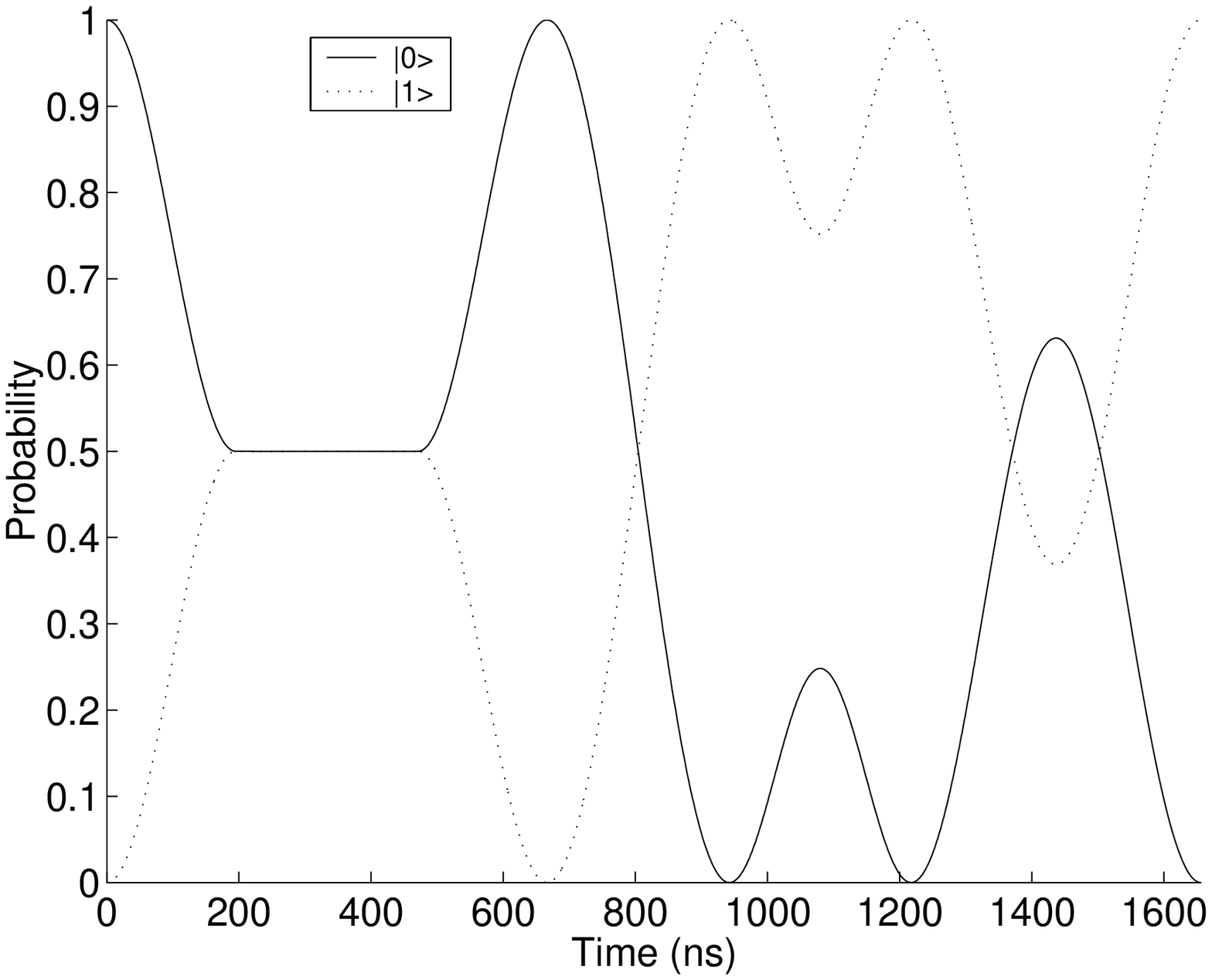}

\caption{Typical Y gate evolution and timescales for input states
as indiacted.\label{fig:YSimulation}}
\end{figure}

For typical parameters expected for the Kane architecture a rotation
of $R_{y}(\pi)$ will take a total time of $\mathrm{89.0\ ns}$ with
the correction step, of which $\mathrm{50.7\  ns}$ is to create
the $Y$ gate, and the remaining $\mathrm{38.3\, ns}$ is used to
correct the rotation of the target qubit with respect to every other
qubit. A typical evolution was numerically simulated and is shown
in Fig. \ref{fig:YSimulation}.

\subsection{Hadamard Gate and Z Rotations}

\begin{figure}
\includegraphics[%
  width=7cm,
  keepaspectratio]{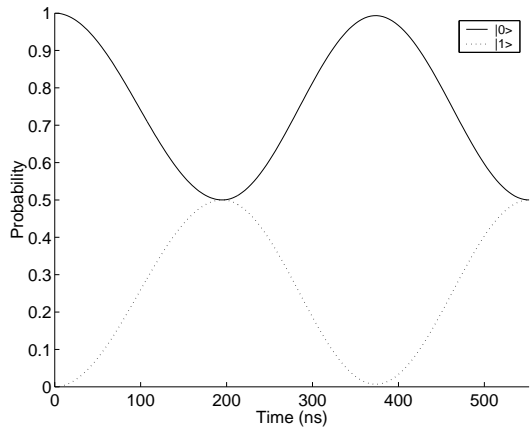}
\caption{Typical evolution and timescales of the Hadamard gate for
states as indicated.\label{fig:HEvolution}}
\end{figure}

Another particularly useful gate in quantum algorithms and quantum
error correction is the Hadamard gate. For reference the Hadamard
gate is defined for a single qubit as:

\begin{eqnarray}
H & = & \frac{1}{\sqrt{2}}\left[\begin{array}{cc}
1 & 1\\
1 & -1\end{array}\right]\nonumber \\
 & = & \frac{1}{\sqrt{2}}(X+Z)\nonumber \\
 & = & R_{m}(\pi)\end{eqnarray}
 where $\hat{m}=\frac{1}{\sqrt{2}}(\hat{\imath}+\hat{k})$. We
may easily produce this gate by detuning the electron spin from
resonance. If we choose $\Delta\omega=\mu_{B}B$ then we will
rotate around the axis $\hat{m}$. Similarly to the X and Y gates,
the Hadamard gate may require a correction step to cancel any
rotation on the spectator qubits. The steps in the Hadamard gate
are showing in Table \ref{tab:HSteps}. The Hadamard gate takes a
total time of $\mathrm{29.7\  ns}$ with the correction applied to
the spectators, but a total of only $\mathrm{10.5\  ns}$ without
the correction step. The Hadamard gate was simulated numerically,
and a typical evolution for this gate is shown in Fig.
\ref{fig:HEvolution}.

\begin{table}
\begin{tabular}{|c|c|c|c|}
\hline Step& Target Qubit& Spectator Qubits& Time
(ns)\tabularnewline \hline \hline 1& $H$& $R_{x}(\alpha)$&
10.5\tabularnewline \hline 2& $I$& $R_{x}(-\alpha)$&
19.2\tabularnewline \hline Overall& $H$& $I$& 29.7\tabularnewline
\hline
\end{tabular}
\caption{Control steps in the single qubit Hadamard gate showing
the operations effected on both target and spectator
qubits.\label{tab:HSteps}}
\end{table}

\begin{figure}
\includegraphics[%
  width=7cm,
  keepaspectratio]{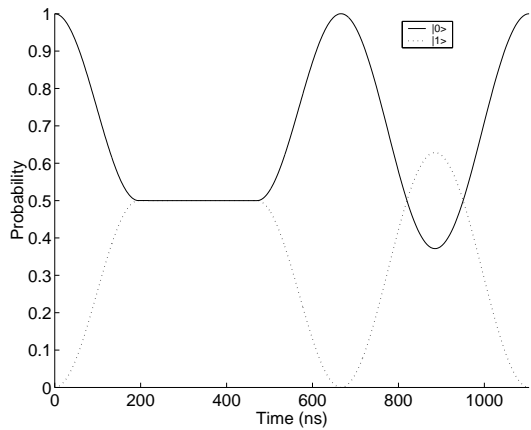}

\caption{Typical evolution and timescales of the Z gate for states as
indicated.\label{fig:ZEvolution}}
\end{figure}

We may perform an arbitrary Z rotation by noting the identity

\begin{equation}
H\  R_{x}(\theta)\  H=R_{z}(\theta).\label{eq:Rz}\end{equation}
Therefore we can simply make an arbitrary Z rotation out of existing
elements. The steps for this gate are shown in Table \ref{tab:ZSteps}.
Only one correction step (Step 4) needs to be applied. The total time
required for this gate is $\mathrm{59.4\  ns}$ with the correction
step included, and $\mathrm{35.8\  ns}$ without the correction step.
Again, a typical evolution is shown in Fig. \ref{fig:ZEvolution}.

\begin{table}
\begin{tabular}{|c|c|c|c|}
\hline Step& Target Qubit& Spectator Qubits& Time
(ns)\tabularnewline \hline \hline 1& $H$& $R_{x}(\alpha)$&
10.5\tabularnewline \hline 2& $R_{x}(\theta)$& $R_{x}(\theta)$&
14.8\tabularnewline \hline 3& $H$& $R_{x}(\alpha)$&
10.5\tabularnewline \hline 4& $I$& $R_{x}(-\theta-2\alpha)$&
23.5\tabularnewline \hline Overall& $R_{z}(\theta)$& $I$&
59.4\tabularnewline \hline
\end{tabular}

\caption{Control steps in the single qubit Z rotation showing the
operations effected on both target and spectator qubits.
\label{tab:ZSteps}}
\end{table}

\section{Multiple Qubit Operations}

\subsection{Exchange Interaction Based CNOT Gate\label{sub:JCNOT}}

The exchange interaction and single qubit unitaries may be used to
create a CNOT gate. The exchange interaction is proportional to
the overlap of the electron wavefunctions. A simple approximation
of the exchange interaction adequate for our purposes, is given
by the Herring-Flicker approximation \cite{Kan98},
\begin{equation}
J_{\max}(d)=\frac{1.6}{\hbar\epsilon}\ \frac{e^{2}}{a^{\star}}
\left(\frac{d}{a^{\star}}\right)^{\frac{5}{2}}
\exp\left(-2\frac{d}{a^{\star}}\right)\label{eqn:Jmax}
\end{equation}
 where $a^{\star}$ is the effective Bohr radius for the electron,
and $d$ is the separation between phosphorus donors. By changing
the voltage of the J gate between the phosphorus donors we may
tune the strength of the exchange interaction $J$ as shown in
\cite{WHK+04,WHP+03,KGS+04}. Ideally the architecture will
be able to tune between $J=0$ and $J=J_{\max}(d)$.

In the rotating frame, the Hamiltonian which includes the exchange
interaction is
\begin{equation}
\tilde{H}_{J}=\mu_{B}B_{ac}(\sigma_{x}^{e_{1}}+\sigma_{x}^{e_{2}})+\Delta\omega_{1}\sigma_{z}^{e_{1}}+\Delta\omega_{2}\sigma_{z}^{e_{2}}+J\bm{\sigma}^{e_{1}}\cdot\bm{\sigma}^{e_{2}},\label{eq:H2Qubit}
\end{equation}
 which is particularly simple to manipulate. Note that if
 $\Delta\omega_{1}=\Delta\omega_{2}$, and in particular if both qubits
 are tuned to the resonant magnetic field meaning
 $\Delta\omega_{1}=\Delta\omega_{2}=0$ the identical single qubit
 rotations commute with the exchange interaction.  That is,
\begin{equation}
\left[\mu_{B}B_{ac}(\sigma_{x}^{e_{1}}+\sigma_{x}^{e_{2}})+\Delta\omega(\sigma_{z}^{e_{1}}+\sigma_{z}^{e_{2}}),\,
  J\bm{\sigma}_{1}\cdot\bm{\sigma}_{2}\right]=0.\label{eq:dotCommutes}
\end{equation}
This implies that we may treat the global rotations and the exchange
interactions separately.

The controlled sign gate $\Lambda_{1}Z$ may be expressed as\begin{equation}
\Lambda_{1}Z=\exp\left(i\pi\frac{I-Z}{2}\otimes\frac{I-Z}{2}\right)\label{eq:CZ}\end{equation}
 Using Hadamard gates, the CNOT gate may be expressed as
 \begin{eqnarray}
\Lambda_{1}X & = & (I\otimes H)\,\Lambda_{1}Z\, (I\otimes H)\nonumber \\
 & = & (H\otimes I)\,\exp\left(i\pi\frac{I-X}{2}\otimes\frac{I-X}{2}\right)\, (H\otimes I)\nonumber \\
 & = & (H\otimes I)\, \left(R_{x}\left(\frac{\pi}{2}\right)\otimes R_{x}\left(\frac{\pi}{2}\right)\right)\,\exp(i\frac{\pi}{4}X\otimes X)\,
 (H\otimes I).\nonumber\\
 \label{eq:CNOT}\end{eqnarray}
Eq. (\ref{eq:CNOT}) is an expression for the CNOT gate which is
mostly made up of gates which are straightforward to perform on our
architecture, such as the Hadamard, and global X rotations. The only
difficult part of this gate is the term $\exp(i\frac{\pi}{4}X\otimes
X)$ which may be constructed in the following way\begin{equation}
\exp(i\frac{\pi}{4}X\otimes X)=(X\otimes
I)\,\exp(i\frac{\pi}{8}\bm{\sigma}\cdot\bm{\sigma})\, (X\otimes
I)\,\exp(i\frac{\pi}{8}\bm{\sigma}\cdot\bm{\sigma}).\label{eq:UmixPi8}\end{equation}
In order to create this interaction correctly, we need to let the
qubits interact for a time $t_{J}$ such that $Jt_{J}=\frac{\pi}{8}$.

The largest amount of time in the CNOT gate is in the correction
operation. In this step, as in previous gates, we rotate the target
qubits with respect to the spectator qubits, and then until the
spectator qubits have performed a whole $2\pi$
rotation. Unfortunately, for the parameters we have chosen, this step
turns out to be particularly long. In the absence of this step, the
CNOT gate requires only an operation time of $96.5 \mathrm{ns}$.

\begin{table}
\begin{tabular}{|c|c|c|}
\hline
Step&
Operation&
Time (ns)\tabularnewline
\hline
\hline
1&
$H\otimes I$&
29.7\tabularnewline
\hline
2&
$\exp i\frac{\pi}{8}\sigma_{1}\cdot\sigma_{2}$&
0.01\tabularnewline
\hline
3&
$X\otimes I$&
14.8\tabularnewline
\hline
4&
$\exp i\frac{\pi}{8}\sigma_{1}\cdot\sigma_{2}$&
0.01\tabularnewline
\hline
5&
$X\otimes I$&
14.8\tabularnewline
\hline
6&
$R_{x}\left(\frac{\pi}{2}\right)\otimes R_{x}\left(\frac{\pi}{2}\right)$&
7.4\tabularnewline
\hline
7&
$H\otimes I$&
29.7\tabularnewline
\hline
8 &
Correction &
51.9\tabularnewline
\hline
Overall&
CNOT&
148.4\tabularnewline
\hline
\end{tabular}

\caption{Control steps and times in the exchange based CNOT
gate.\label{tab:CNOTTimes}}
\end{table}

\begin{figure}
\includegraphics[%
  width=7cm,
  keepaspectratio]{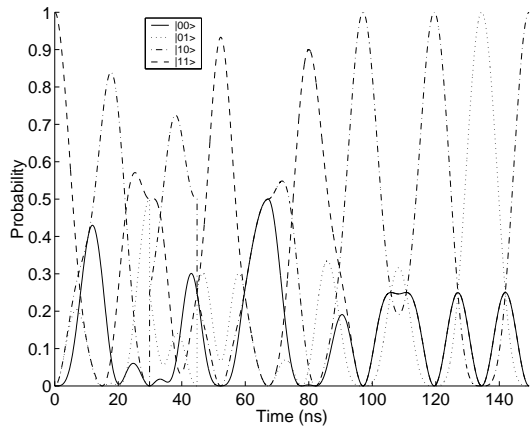}
\caption{Evolution and timescales for the CNOT gate for the states as
indicated.\label{fig:CNOTEvolution}}
\end{figure}

\begin{figure*}
\includegraphics[%
  width=10cm,
  keepaspectratio]{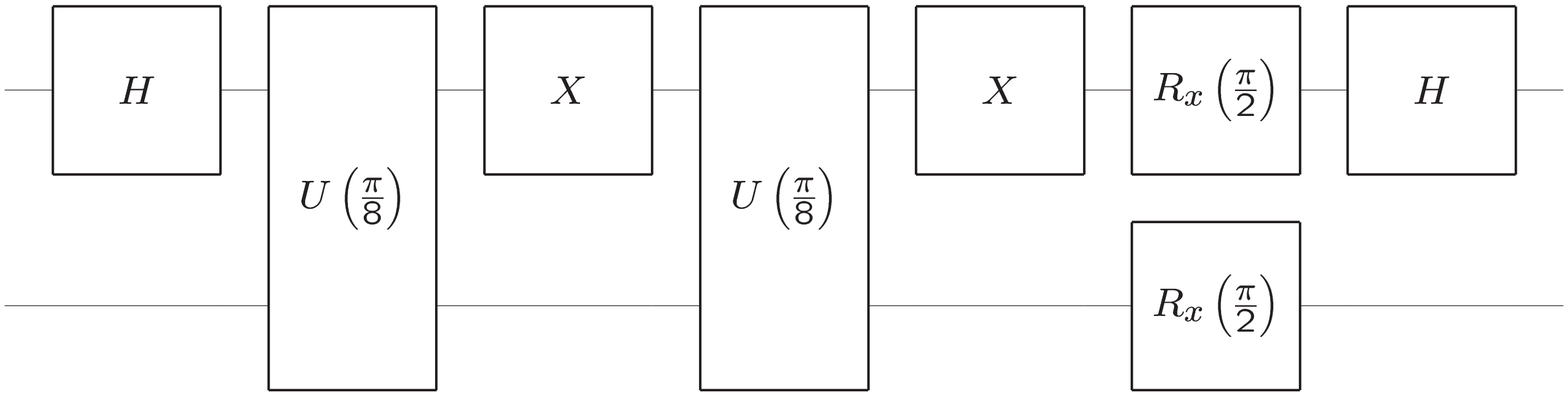}
\caption{Circuit diagram for the CNOT gate.\label{fig:CNOTCircuit}}
\end{figure*}

The circuit diagram based on this construction is shown in Fig.
\ref{fig:CNOTCircuit}.  The total time required for this gate, based
on typical parameters for the Kane architecture is $\mathrm{148.4\
ns}$. A breakdown of the times required for each operation in the gate
is shown in Table \ref{tab:CNOTTimes}. This gate was simulated
numerically, and a typical simulation is shown in Fig.
\ref{fig:CNOTEvolution}. Note that during this gate, corrections need
to be performed only when they do not commute with the next gate.

\subsection{The Swap Gate}

The swap gate may be performed particularly easily with the exchange
interaction. The swap gate $S$ may be written
\begin{equation}
S=\exp(i\frac{\pi}{4}\bm{\sigma}_{1}\cdot\bm{\sigma}_{2}).\label{eq:Swap}
\end{equation}
Assuming control of the exchange interaction, this gate may be
performed in a single operation with $Jt=\frac{\pi}{4}$. This may then
require a correction step. Since this interaction is much larger than
the typical frequencies for single qubit rotations, this gate is
extremely fast, and to a good approximation does not require a
correction step. The speed of this gate also indicates that a three
qubit encoding \cite{DBK+00} may be successful.

\subsection{The Dipole-Dipole Based CNOT Gate} \label{sub:DCNOT}

The dipole-dipole interaction couples every pair of electronic spins
in the system. The contribution which the dipole-dipole interaction
makes to the Hamiltonian is

\begin{equation}
H_{D}=D\left(\bm{\sigma}^{e_{1}}\cdot\bm{\sigma}^{e_{2}}-3(\bm{\sigma}^{e_{1}}\cdot\hat{d})\otimes(\bm{\sigma}^{e_{2}}\cdot\hat{d})\right)\label{eq:HD}
\end{equation}
where the strength of the dipole-dipole interaction $D$ is given
by\begin{equation}
D(d)=\frac{\mu_{0}}{4\pi}\frac{\mu_{B}^{2}}{d^{3}}.\label{eq:D}
\end{equation}
Whereas the exchange interaction dies off exponentially with distance,
as shown in Eqn. {(}\ref{eqn:Jmax}{)}, the dipole-dipole interaction
only dies off as $1/d^{3}$. Therefore at larger separations, the
dipole-dipole interaction to dominates.

The direction in which we orientate our qubits w.r.t. the magnetic
field $B$ is important. If we align the donors along the x-axis
($\hat{\imath}$) or y-axis ($\hat{\jmath}$), $H_{D}$ does not commute
with $\sigma_{z}^{e_{1}}+\sigma_{z}^{e_{2}}$.  This implies we may no
longer look at our system in a rotating frame.  However, if we align
our qubits in the z-axis ($\hat{k}$) direction then the rotating frame
is still valid, and the rotating frame Hamiltonian is
\begin{eqnarray}
\tilde{H}_{JD} & = &
\mu_{B}B_{ac}(\sigma_{x}^{e_{1}}+\sigma_{x}^{e_{2}})+\Delta\omega_{1}\sigma_{z}^{e_{1}}+\Delta\omega_{2}\sigma_{z}^{e_{2}}\nonumber
\\ & &
+J\bm{\sigma}^{e_{1}}\cdot\bm{\sigma}^{e_{2}}+D\left(\bm{\sigma}^{e_{1}}\cdot\bm{\sigma}^{e_{2}}-3\sigma_{z}^{e_{1}}\otimes\sigma_{z}^{e_{2}}\right).\label{eq:HJD}\end{eqnarray}
For simplicity we therefore choose to align our axes in the $\hat{k}$
direction.

When the electrons are relatively widely spaced ($d>\mathrm{{30\,
nm}}$) single qubit rotations are much faster than the speed of the
interaction ($\mu_{B}B_{ac}\gg D$) and the dipole-dipole interaction
dominates the exchange interaction ($D\gg J$). In order to specify a
CNOT gate, we consider the case when the electrons are tuned to the
rotating magnetic field, i.e. $\Delta\omega_{1}=\Delta\omega_{2}=0$.

Now
\begin{equation}
[\mu_{B}B_{ac}(\sigma_{x}^{e_{1}}+\sigma_{x}^{e_{2}}),\,(J+D)\bm{\sigma}^{e_{1}}\cdot\bm{\sigma}^{e_{2}}]=0\label{eq:DotJDCommute}
\end{equation}
therefore we consider the interaction $(J+D)\bm{\sigma}^{e_{1}}\cdot\bm{\sigma}^{e_{2}}$
separately from the single qubit rotations $\mu_{B}B_{ac}(\sigma_{x}^{e_{1}}+\sigma_{x}^{e_{2}})$.
Unfortunately the same is not true of the $-3D\sigma_{z}^{e_{1}}\otimes\sigma_{z}^{e_{2}}$
term in the Hamiltonian where
\begin{equation}
[\sigma_{z}^{e_{1}}\otimes\sigma_{z}^{e_{2}},\,\sigma_{x}^{e_{1}}+\sigma_{x}^{e_{2}}]=2i(\sigma_{y}^{e_{1}}\otimes\sigma_{z}^{e_{2}}+\sigma_{z}^{e_{1}}\otimes\sigma_{y}^{e_{2}}),\label{eq:ZXDontCommute}
\end{equation}
for example. Similarly we may calculate higher order commutators.
This leads to quite a complicated evolution of the system. Fortunately
it is possible to refocus \cite{Sli78} much of the evolution. However,
these higher order terms also anti-commute with
$\sigma_{x}^{e_{1}}\otimes I$ and therefore we may cancel many of them
by conjugation. With this approximation, it is possible to create the
CNOT gate using exactly the same pulse sequence as was required when
we ignored the dipole-dipole interaction in
Section\ref{sub:JCNOT}. The circuit diagram for this circuit is shown
in Fig. \ref{fig:CNOTCircuit}. The interaction is now assumed to be
solely due to weak dipole-dipole interaction. In each interaction we
must allow the qubits to interact for the comparatively long time of
$t_{D}=\frac{1}{D}\frac{\pi}{8}$.

At a spacing of $\mathrm{30\, nm}$ we anticipate an extremely long
gate time of $\mathrm{4.6\, ms}$. This time is dominated by the time
required for the interaction between qubits. A quantum computer based
on this scheme has no need for J gates.

\subsection{CNOT Gate with Both Exchange and Dipole-Dipole Interactions}

In the intermediately spaced regions, neither the dipole-dipole nor
the exchange interaction dominate. In this case we may use them to
complement each other, and create a CNOT gate as described in Section
\ref{sub:JCNOT} and Section \ref{sub:DCNOT}. In this case the
interaction between electrons must be performed for a time of
$t_{JD}=\frac{1}{J+D}\frac{\pi}{8}$.  This leads to a total gate time
(for typical parameters at a spacing of $\mathrm{23\, \rm{nm}}$) of
$\mathrm{4.0\,\mu s}$.

During single qubit rotations it would be beneficial (although not
essential) to minimise the exchange interaction. This may be accomplished
through the application of voltage to the J gates to isolate the electrons.

\section{Parallel gate operation}

Parallel gates are an essential feature of scalable architectures, and
are performed naturally in this scheme.  Each of the gates may be
performed in parallel.  For example X rotations may be performed on
two qubits at the same time. This is achieved by simply applying
identical control pulses to both qubits. Similarly, identical
two-qubit operations may be performed in parallel. For example, a CNOT
may be performed between qubits one and two, and between three and
four, in parallel.

In addition to applying identical gates to different qubits, many
other combinations are possible. Every gate takes a multiple of the
period of a spectator qubit ($29.7\,\mathrm{ns}$) to perform.  After a
whole number of periods, the spectators will be in their original
orientation. During this time, the correct rotation is applied to the
target qubits. The shorter operation being applied in parallel may
have to be padded by a number of $2\pi$ X rotations exactly the same
way as the spectators. In this way any two operations which do not act
on the same qubits may be applied in parallel. So, for example, an X
rotation on qubit 1, may be performed in parallel with a CNOT on
qubits 2 and 3.

Our scheme takes advantage of two key facts. Firstly, each gate only
requires us to change the voltage on the local `A' and `J' gates. We
do not need to modify the magnetic fields, which would affect the
operation of other qubits. This means that each operation may be
applied independantly. Secondly, each operation takes a whole number
of periods of the spectator qubits to perform. Much like a clock in a
conventional computer, this greatly simplifies timing issues in
performing gates in parallel.

\section{Readout and Initialisation}

Readout is a crucial issue to be addressed for donor spin based
architectures. We will briefly describe several possible readout
schemes here.

Direct single-spin detection is very difficult since a single spin
interacts very weakly with its environment and hence the measurement
device. In spite of this, magnetic resonance force microscopy (MRFM)
has been suggested \cite{Sid91, Sid92, BBG+03, BG03} as one of the
most promising techniques to achieve such a direct single-spin
measurement. Two of the most promising spin-cantilever modulation
protocols to detect a single spin by MRFM are: Cyclic Adiabatic
Inversion (CAI) \cite{WBY+98} and OScillating Cantilever-driven
Adiabatic Reversal (OSCAR) \cite{MBC+03}. The MRFM technique also
takes the advantage of the electron spin quantum computer architecture
discussed here.  The required RF field for the MRFM measurement
protocols is also an essential element for the electron spin quantum
gate operations. Recently, the MRFM technique has been demonstrated
\cite{RBM+04} to detect an individual electron spin. But the required
averaging time is still too long to achieve the real-time readout of
the single electron spin quantum state. Given the steady improvement
in experimental technique, the MRFM has great potential to serves as
an readout device for spin-based qubit systems in the near future.

\begin{figure}
\includegraphics[%
  width=7cm,
  keepaspectratio]{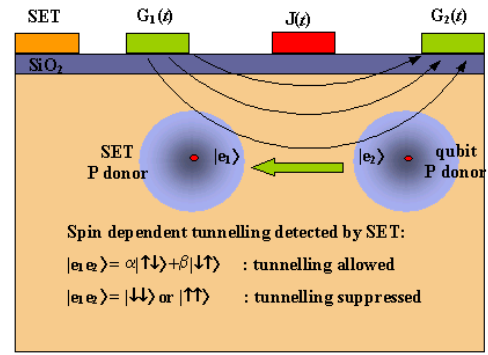}
\caption{Schematic of the spin-charge transduction process for
spin readout using a single electron transistor (SET) as an
electrometer.\label{fig:KaneReadout}}
\end{figure}

The spin-charge transduction idea of the original Kane proposal
(Fig. \ref{fig:KaneReadout}) called for the adiabatic spin-dependent
transfer\cite{Kan98} of the qubit donor electron to an auxiliary donor
leaving a donor ion D$^+$ and a doubly occupied donor D$^-$. The two
electron state of this double donor system is conditionally entangled
with the original nuclear qubit spin.  Detection of the final
D$^+$D$^-$ state by the SET constitutes a measurement of the qubit
nuclear spin. A problem with this scheme is the shallow nature of the
D$^-$ state (1.7 meV), which may easily ionise in the electric field
required to induce the electron transfer.

The dynamics of the spin dependent transition $D^0D^0\rightarrow
D^+D^-$ was investigated to assess the vulnerability of the adiabatic
read-out scheme\cite{HWP+04}. A comparison to the field strength
required for adiabatic transfer, the typical D$^-$ state dwell-times
and SET timescales indicated that adiabatic transfer would at the
every least severely test SET measurement capability.  As a possible
alternative to the adiabatic Kane proposal, a resonant-based
\cite{HWP+04} scheme has been proposed in which an AC field is applied
to the gates $G_{1,2}$ resonant with the transition $D^0D^0\rightarrow
D^+D^-$. Simulation results indicate a good level of control is
achievable for single-qubit addressing in this way using relatively
low DC field strengths.

Another alternative\cite{GHH+04} relies on energy resolved readout
through the introduction of an ionised donor (the probe) to the
usual two donor system for spin readout. Controlling the bias
applied to the probe allows resonant charge transfer from either
the singlet or triplet state of the combined qubit-reference
system to the probe. By effecting spin dependent tunnelling to the
ionised probe, rather than to the reference in the two-donor
scheme, we avoid potential problems due to shallow the $D^-$
state.  This can be thought of as using a charge qubit to readout
a spin qubit.

\section{Conclusion}

We have proposed a scheme for solid state quantum computation based on
donor electron spins and global control, using only weak local
control. This scheme forms a natural stepping stone and shares
similarities with the existing nuclear spin based Kane proposal. We
have shown how, even with limited control over the resonant
frequencies of the electronic spins and an always on rotating magnetic
field, $B_{ac}$, this system may be used for quantum computation. This
scheme outperforms the na\"ive application of the canonical scheme.
Indeed, although electron dephasing times are faster than the
corresponding nuclear dephasing times, we find that a typical
operation time is also correspondingly faster with with
$\mathrm{T_{2}/T_{ops}}$ approaching $10^{6}$.

\section{Acknowledgements}

CDH and HSG would like to thank Gerard Milburn for support. This
work was partly supported by the Australian Research Council and
by the US National Security Agency (NSA), Advanced Research and
Development Activity (ARDA) and the Army Research Office (ARO)
under contract number DAAD19-01-1-0653. H.S.G. would like to
acknowledge financial support from Hewlett-Packard.

\bibliography{bibliography}

\end{document}